\documentclass[twoside,web]{ieeecolor}
\usepackage{generic}
\usepackage{cite}
\usepackage{hyperref}
\usepackage{amsmath,amssymb,amsfonts}
\usepackage{algorithmic}
\usepackage{graphicx}
\usepackage{textcomp}

\def\BibTeX{{\rm B\kern-.05em{\sc i\kern-.025em b}\kern-.08em
    T\kern-.1667em\lower.7ex\hbox{E}\kern-.125emX}}
\begin{document}

\title{{\color{red}In-Memory AM Demodulation Using an All-Silicon Independent-Dual-Gate Gain-Cell Memory with $<10^{-22}$ A Leakage Determined by Single-Electron Counting}}
\author{Katsuhiko Nishiguchi,\IEEEmembership { Member, IEEE}, Toshiaki Hayashi, Kensaku Chida, Takase Shimizu, Gento Yamahata, and Seiya Kasai,\IEEEmembership { Senior Member, IEEE}
\thanks{This work was supported by JSPS KAKENHI Grant Number JP26H02150.}
\thanks{K. Nishiguchi is with Institute for Multidisciplinary Sciences, Yokohama National University, Kanagawa 240-8501, Japan (e-mail: nishiguchi-katsuhiko-kr@ynu.ac.jp).}
\thanks{T. Hayashi, K. Chida, T. Shimizu, and G. Yamahata are with Basic Research Laboratories, NTT Inc., Kanagawa 243-0198, Japan.}
\thanks{S. Kasai is with Research Center for Integrated Quantum Electronics, Hokkaido University, Hokkaido 060-0813, Japan.}
\thanks{\copyright\ 2026 IEEE. Personal use of this material is permitted. Permission from IEEE must be obtained for all other uses, in any current or future media, including reprinting/republishing this material for advertising or promotional purposes, creating new collective works, for resale or redistribution to servers or lists, or reuse of any copyrighted component of this work in other works.}}

\maketitle

\begin{abstract}
Ultra-low-leakage memories are attracting increasing attention for in-memory sensing and computing. However, achieving sufficiently long retention in silicon memories for analog signal processing remains challenging because of leakage through the access transistor. In this work, we demonstrate an all-silicon independent-dual-gate memory whose leakage current, inferred from single-electron counting statistics, is below $10^{-22}$ A, giving a measured retention time exceeding 1000 s. Owing to the extremely low leakage, the subthreshold nonlinearity of the access transistor can be exploited without disturbing the stored charge, enabling in-memory amplitude-modulation (AM) demodulation. The proposed memory provides a CMOS-compatible platform for ultra-low-power signal processing and in-memory sensing. 
\end{abstract}


\section{Introduction}
\label{sec:introduction}
In-memory computing and in-memory sensing have attracted increasing attention \cite{Sun2023,Fantini2025, Zhou2020} because they reduce data transfer between memories and processors, leading to low-power and high-speed signal processing. In particular, direct analog signal processing inside memories is expected to become an important function for future edge sensing systems. Such operation, however, requires memory cells that can retain analog information for sufficiently long periods while allowing repeated readout without disturbing the stored charge.

Oxide-semiconductor memories have recently demonstrated long-retention characteristics suitable for analog information storage \cite{Oota2019,Belmonte2020,Belmonte2021,Fujii2024,Li2025}. However, their compatibility with mainstream CMOS technology is still unclear. Silicon memories are highly attractive because of their mature fabrication technology and high scalability, but their retention characteristics are generally limited by leakage through the access transistor, such as gate-induced drain leakage (GIDL) \cite{Chan1987}, preventing long-term analog information storage.

In this work, we exploit an independent-dual-gate gain-cell memory (iDGCM) in which the storage node is electrically isolated from leakage paths. The leakage current is { inferred to be below $10^{-22}$ A, enabling a retention time exceeding 1000 s}. This ultra-low leakage enables in-memory AM demodulation based on the intrinsic nonlinear subthreshold characteristics of the access transistor. While the conversion of an alternating-current (AC) signal to  a direct-current (DC) signal has been reported in a single-electron DRAM \cite{Salhani2025}, AM demodulation requiring selective frequency discrimination via a tunable cut-off frequency has not been demonstrated. This work establishes ultra-low leakage as the key enabler for extending in-memory operation beyond storage toward analog signal processing.

\section{Device Structure}

\begin{figure}[!t]
\centerline{\includegraphics[width=\columnwidth]{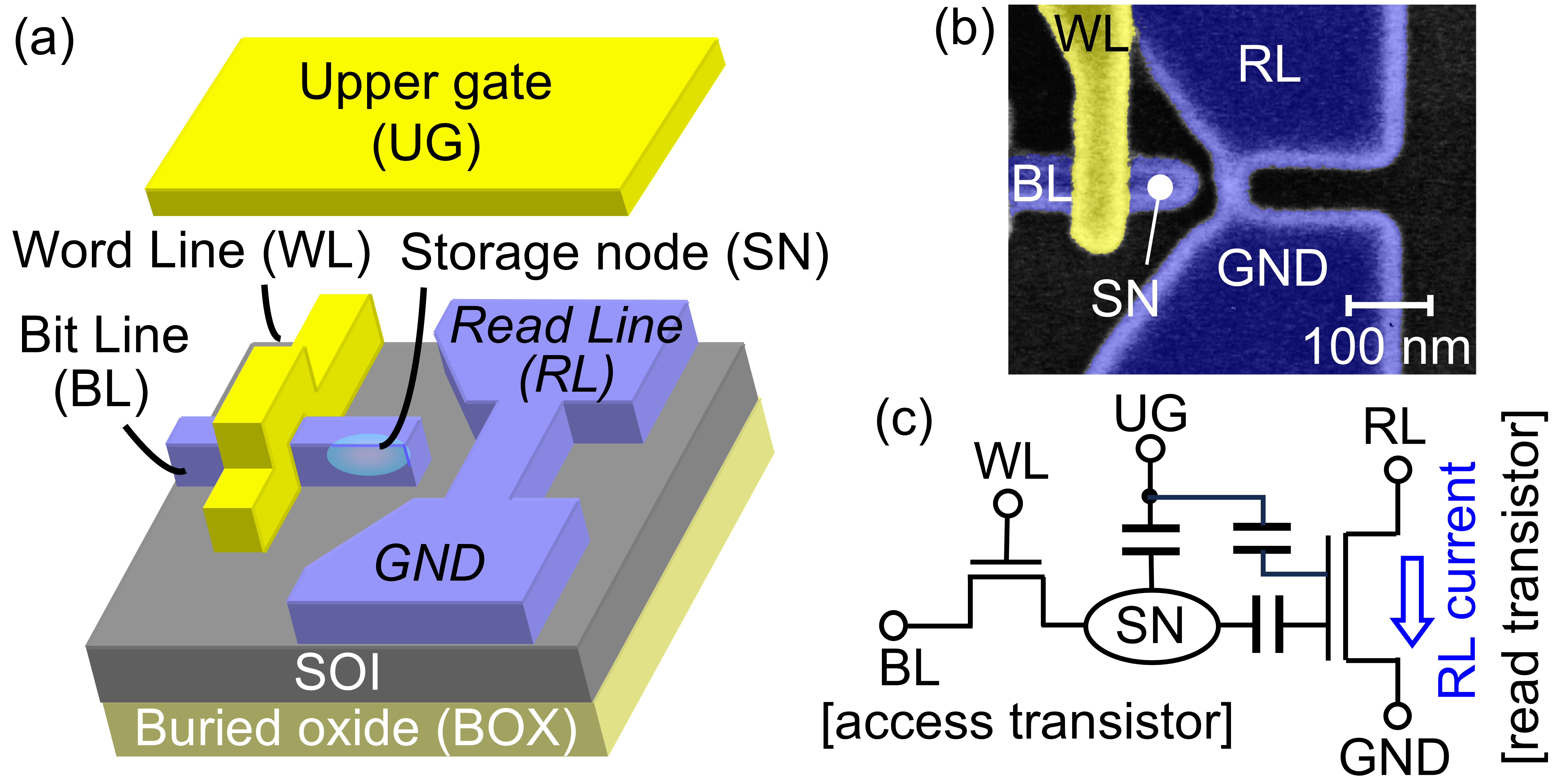}}
\caption{(a) Bird's-eye view and (b) false-color SEM image of the device fabricated on a silicon-on-insulator (SOI) wafer. The word line (WL) and upper gate (UG) are made of poly-Si. After the fabrication step shown in (b), the device is oxidized. {The gate-oxide thickness between the SOI channel and WL is about 40 nm.} The entire area shown in (b) is covered by a 50-nm-thick interlayer $\rm SiO_2$ formed by a chemical vapor deposition, followed by the UG}. After its formation, phosphorus ions are implanted into the SOI channel to form terminals connected to the bit line (BL), read line (RL), and ground (GND). The thickness and width of a silicon-on-insulator channel are 10 nm each. The gate length is about 10 nm. The oxide thickness between the WL and UG is about 70 nm. The buried-oxide thickness is 400 nm.
(c) Equivalent circuit. The RL voltage was fixed at 500 mV throughout this paper.
\label{fig1}
\end{figure}

The iDGCM is fabricated on a silicon-on-insulator wafer (Fig. 1) \cite{Nishiguchi2007}. The attofarad-scale storage node (SN) stores charge transferred from the bit line (BL) via the word line (WL) of an access transistor. A key feature is its intrinsic Si channel, independently controlled by two gates. The upper gate (UG), held at constant bias, forms an inversion layer in the channel, effectively serving as a drain connected to the BL. The WL (10 nm gate length) acts as a transfer gate injecting charge into the SN. This dual-gate structure allows the removal of a pn junction beneath the WL—a primary source of GIDL and trap-assisted tunneling leakage \cite{Hamamoto1998}—enabling ultra-long retention. As shown in Fig. 2, a reference transistor with the double-gate structure exhibits an off-current below $10^{-14}$ A. Additionally, the dual gates offer distinct roles: the WL defines the effective channel length, and its threshold voltage is tunable via the UG bias. 
Despite capacitive coupling between the WL and UG, charge transfer in a Si transistor of similar geometry has been reported up to 6.5 GHz \cite{Yamahata2016}, suggesting that this coupling is unlikely to limit charge transfer in this frequency range.  
All the results presented below were obtained from a single iDGCM; device-to-device variability remains to be characterized in future work.

\begin{figure}[!t]
\centerline{\includegraphics[width=\columnwidth]{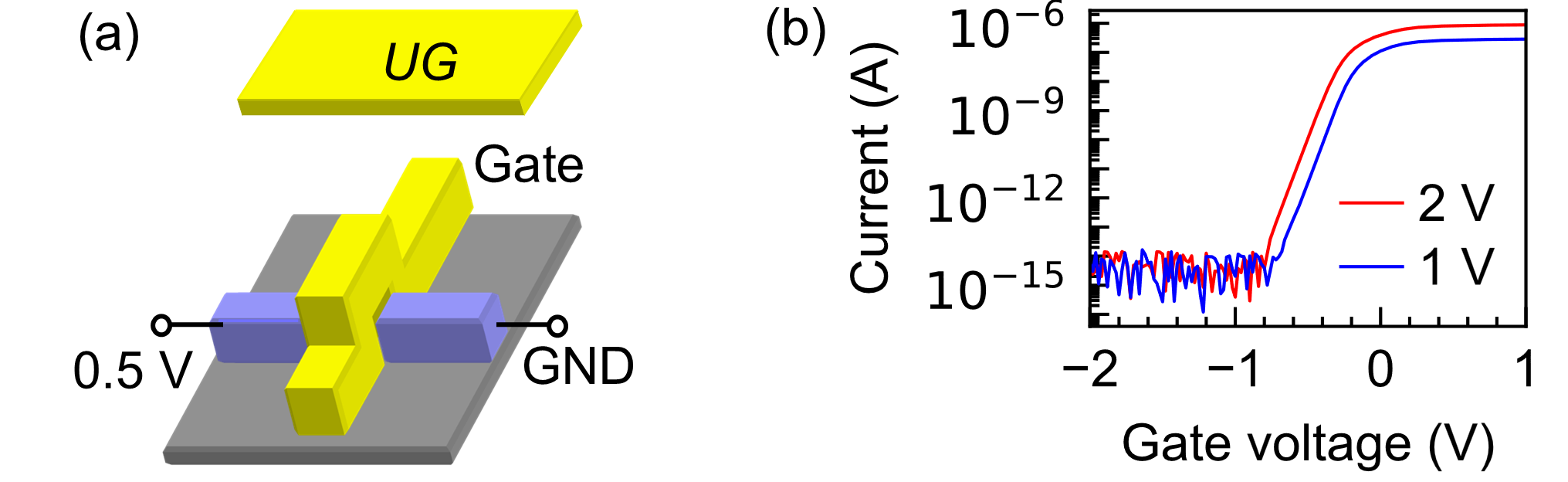}}
\caption{ Characterization of a reference transistor with the same channel and gate geometry as the access transistor comprising the iDGCM. (a) Schematic of the reference transistor. (b) Its measured transfer characteristics at UG voltages of 1 and 2 V. The current below $10^{-14}$ A reflects the measurement noise floor of the system. }
\label{fig2}
\end{figure}

\section{Experimental Results at room temperature (298 K)}

\subsection{Retention characteristics}

Figure 3 shows the retention characteristics of the iDGCM after data '1' and '0' were stored. Data ‘1’ remains clearly separated from data ‘0’ even after 1000 s [see Fig. 3(b)]. A downward step $\Delta I_e$ after a time interval of $\Delta t$ directly reflects a single-electron leakage from the BL to the SN \cite{Nishiguchi2008}: The leakage current is estimated from $e/\Delta t_{\rm avg}$, where $e$ is the elementary charge, and $\Delta t_{\rm avg}$ is the average value of $\Delta t$ obtained from repeated writings of data '1'. As shown in Fig. 3(c), the leakage current exhibits a subthreshold exponential dependence on the WL voltage $V_{\rm off}$, falling below $10^{-22}$ A and decreasing further with more negative values of $V_{\rm off}$. 
These characteristics imply that leakage mechanisms, such as GIDL and defect-assisted current, are suppressed at least down to $10^{-22}$ A. 
Note that this value is not a DC current reading but a statistical inference from single-electron counting. This circumvents the noise floor of conventional current measurement, about $10^{-14}$ A for our setup [Fig. 2(b)], eight orders of magnitude higher.

In this study, the small SN capacitance $C$, estimated to be 7.7 aF from $\Delta I_e$ and transconductance of the read transistor \cite{Nishiguchi2008}, resolves single-electron leakage events. Since ultra-low leakage is independent of SN capacitance, the concept is applicable to larger memories with longer retention.  

\begin{figure}[!t]
\centerline{\includegraphics[width=\columnwidth]{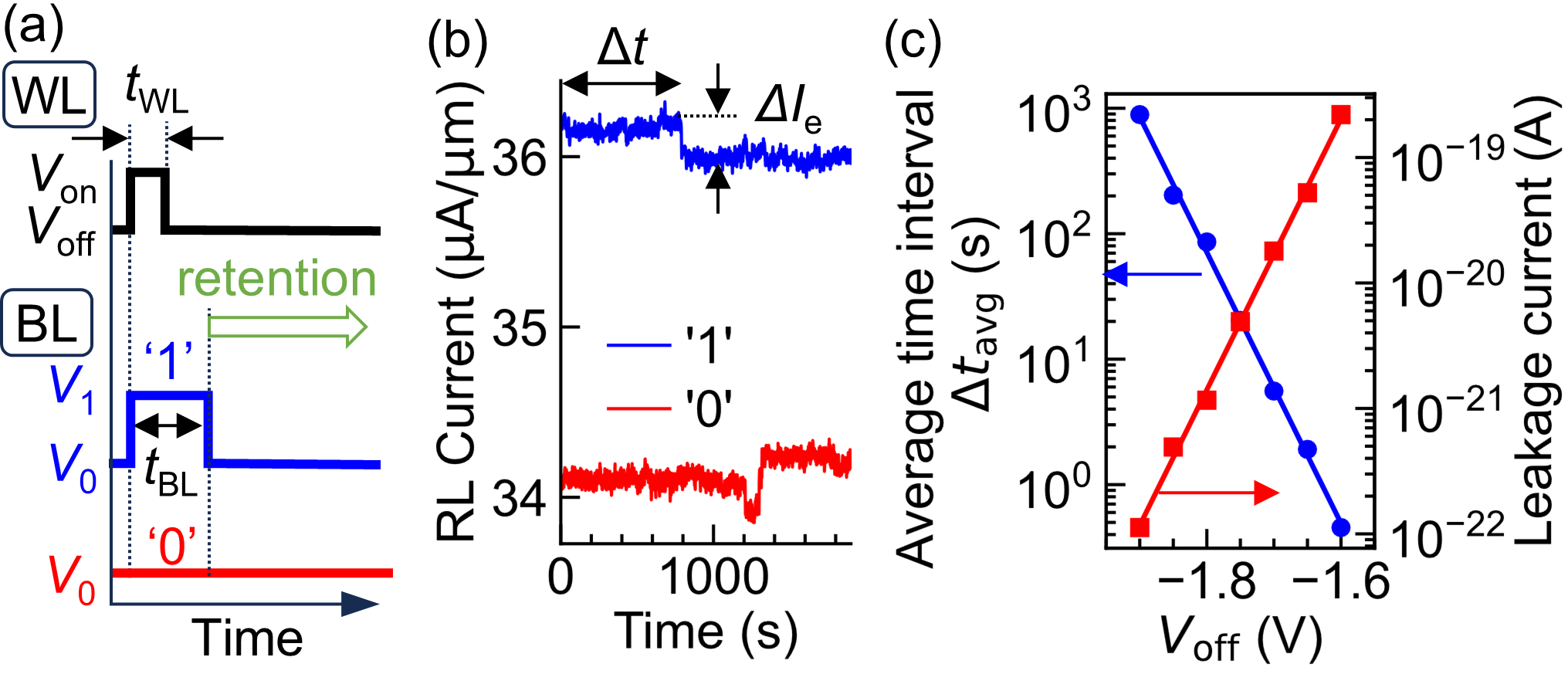}}
\caption{(a) Timing sequence for writing ‘1’ and ‘0’ and evaluating retention. $V_{\rm on}$, $V_{\rm 1}$, and $V_{\rm 0}$ were 0.2, 1, and 0 V, respectively. $t_{\rm WL}$ and $t_{\rm BL}$ were 20 and 40 ns, respectively. (b) Retention characteristics measured via the RL current. $V_{\rm off}$ was -1.9 V. $\Delta I_e$ denotes the current change associated with a single electron entering the SN, and the corresponding time interval is defined as $\Delta t$. 
(c) Measured average time interval $\Delta t_{\rm avg}$ for one electron entering the SN as a function of $V_{\rm off}$. The symbols are experimental data, and the solid lines are exponential fits to them.
$\Delta t_{\rm avg}$ is obtained by averaging $\Delta t$ from 100 repeated writings of data ‘1'. Leakage current given by $e/\Delta t_{\rm avg}$ is also shown on the right axis.}
\label{fig3}
\end{figure}

\subsection{In-Memory Demodulation for AM Signals}

The wide tunability of $\Delta t_{\rm avg}$ of the low-leakage access transistor enables in-memory AM demodulation. This function is based on AC-to-DC conversion \cite{Salhani2025}: An AC signal, $S_{\rm amp}{\rm sin}(2\pi ft)$, is applied to the BL, and the resulting SN voltage is read out as the RL current. As demonstrated in Fig. 4(a), at 0.1 Hz, the RL current as an output signal follows the sinusoidal input waveform, whereas at 10 MHz, the RL current saturates at a constant value, corresponding to the AC-to-DC conversion. To quantify the AC-to-DC conversion ratio, we evaluate the SN voltage [the right axis in Fig. 4(a)], derived from $e\Delta I_{\rm ac}/C\Delta I_e$, where $\Delta I_{\rm ac}$ is  the difference in current between when an AC signal is applied and when it is not.

\begin{figure}[!t]
\centerline{\includegraphics[width=\columnwidth]{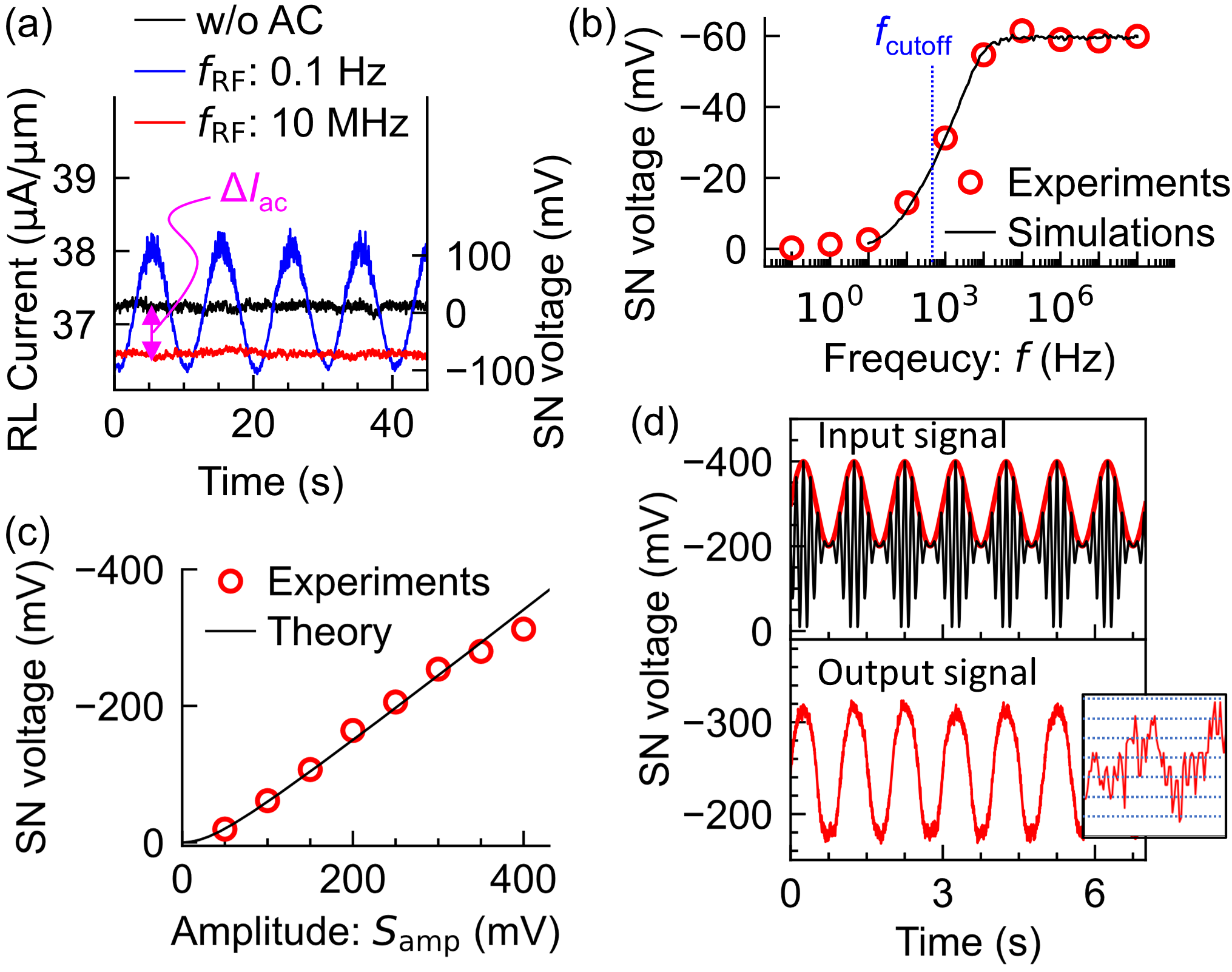}}
\caption{(a) AC-to-DC conversion. A sinusoidal signal, $S_{\rm amp}\rm{sin}(2 \pi ft)$, is superimposed on the BL. The measured RL currents with and without the AC signal are compared (left axis). $S_{\rm amp}$ was 100 mV. In (a)-(c), the WL voltage was -1.3 V. 
The right axis is the corresponding SN voltage converted from the measured RL current using $e\Delta I_{\rm ac}/C\Delta I_e$. $\Delta I_{\rm ac}$ and $\Delta I_e$ are the changes in the RL current caused by the AC signal and by one electron entering the SN [see Fig. 3(b)], respectively. $C$ of 7.7 aF is the experimentally estimated capacitance of the SN.
(b) The SN voltage as a function of frequency $f$ of the AC signal. $S_{\rm amp}$ was 100 mV. The symbols are experimental data converted from the measured RL current, and the solid line is the Monte-Carlo simulation carried out with 100,000 iterations. The dotted line depicts $\Delta t_{\rm avg}^{-1}$ of 500 Hz, which corresponds to the cut-off frequency $f_{\rm cutoff}$ of the access transistor. 
(c) $S_{\rm amp}$ dependence of the SN voltage. $f$ was 100 MHz. The symbols are experimental data. The solid line is the calculated curve given by $(e\beta)^{-1} {\rm ln}[I_0(e\beta S_{\rm amp})]$, where $I_0(\Box)$ is the modified Bessel function of the first kind of the order zero, and $\beta=(k_{\rm B}T)^{-1}$ is the inverse temperature \cite{Salhani2025,Salhani2026}.                             
(d) AM demodulation. The upper figure is a calculated waveform of the AM signal applied to the BL; for clear visibility, the carrier frequency is drawn as 7 Hz instead of the actual 100 MHz. The bold line is the envelope of the signal modulating the carrier signal. The frequencies of the modulating and carrier signals were 1 Hz and 100 MHz, respectively. The lower figure is the output SN voltage converted from the measured RL current. The inset is the enlarged plot of the lower figure. The dotted lines represent the signal levels discretized by single electrons entering and leaving the SN. }
\label{fig4}
\end{figure}

Figure 4(b) shows the $f$ dependence of the time-averaged SN voltage. At $f \ll \Delta t_{\rm avg}^{-1}$ (= 500 Hz), the average SN voltage remains close to zero due to its oscillatory behavior, whereas it saturates at a constant negative value for $f \gg \Delta t_{\rm avg}^{-1}$, indicating AC-to-DC conversion. 
This $\Delta t_{\rm avg}^{-1}$ corresponds to the cut-off frequency $f_{\rm cutoff}$ of the access transistor. In the steady state at a constant BL voltage, electrons shuttle between the BL and the SN with no net flow: they enter the SN with an average interval $\Delta t_{\rm avg}$ set by the energy barrier height in the channel under the WL, and the same number of electrons leaves the SN, so that the average interval for leaving is also $\Delta t_{\rm avg}$. The resulting fluctuation of the SN charge has a Lorentzian spectrum with a corner frequency of $\Delta t_{\rm avg}^{-1}$, which was experimentally confirmed in \cite{Nishiguchi2014}. Since the barrier height does not depend on the stored charge, $ \Delta t_{\rm avg}$ measured in Fig. 3(c) gives $f_{\rm cutoff}$ directly. 
  
Figure 4(c) shows the SN voltage as a function of $S_{\rm amp}$ at $f \gg f_{\rm cutoff}$. When $S_{\rm amp}>2k_{\rm B}T/e \approx$ 50 mV, the SN voltage decreases with increasing $S_{\rm amp}$ with a proportionality close to -1, which is ensured by the theoretical prediction \cite{Salhani2025}. Here, $k_{\rm B}$ is the Boltzmann constant and $T$ is the temperature. This linear relationship enables faithful AM envelope detection without requiring precise gain control or signal calibration. 

In Figs. 4(b) and (c), the theoretical and simulation lines, both of which reproduce the experimental results well, are derived not from fitting parameters but from experimentally obtained values including $C$ and $\Delta t_{\rm avg}$. The theory and simulation demonstrate that the operating principle of AC-to-DC conversion lies in the nonlinear characteristics of the transistor \cite{Salhani2025, Salhani2026}, confirming that the experimental AC-to-DC conversion is achieved by the ultra-low leakage current in the subthreshold current region of the access transistor. 

An in-memory demodulation function for AM signals is implemented using these AC-to-DC conversion functions. The AM signal is generated by superimposing a modulating signal, whose frequency is $f_{\rm mod}$, onto a high-frequency carrier signal with the frequency $f_{\rm carrier}$ as shown in the upper figure of Fig. 4(d). The envelope, highlighted with the thick line, directly represents the modulating signal. 

For demodulation operation, $\Delta t_{\rm avg}$ is adjusted using the WL so that $f_{\rm mod}<f_{\rm cutoff}<f_{\rm carrier}-f_{\rm mod}<f_{\rm carrier}$. Under these conditions, the SN voltage reflects the amplitude of frequency components higher than $f_{\rm cutoff}$, effectively representing the AM envelope. The lower figure of Fig. 4(d) confirms accurate reproduction of the sinusoidal envelope of the AM signal applied to the BL. Although the output signal is noisy, this noise comes from the small SN increasing its $k_{\rm B}T/C$ voltage noise \cite{Nishiguchi2014}. In fact, discrete voltage fluctuation caused by single-electron motion is observed as shown by the inset figure. Since a small SN is not essential for AM demodulation, a larger SN is useful for reducing noise. This in-memory demodulation potentially reduces the need for external analog front-end circuits such as mixers, filters, or amplifiers, supporting analog sensing with minimal overhead.
This overhead can be quantified: the power dissipated during this operation, given by $f_{\rm cutoff}eS_{\rm amp}I_1(\beta e S_{\rm amp})$ \cite{Salhani2026}, is at most $2\times 10^{-11}$ W for the envelope range in Fig. 4(d), where $I_1(\Box)$ is the modified Bessel function of the first kind of order one. This is an intrinsic value that excludes stray capacitances. The total is dominated by the read transistor, which consumes about 0.2 $\mu$W.

The theoretical requirement for $f_{\rm carrier}$ is $f_{\rm carrier} > f_{\rm cutoff}$ so that the average SN voltage reaches its saturated value [Fig. 4(b)]. Demodulation was demonstrated here at $f_{\rm carrier}$ of 100 MHz; operation up to the GHz range is a possible extension rather than a demonstrated capability, since our present setup is not optimized for high-frequency signal transmission. The theory imposes no upper bound on $f_{\rm carrier}$, and charge transfer through a Si transistor of similar geometry has been reported at up to 6.5 GHz \cite{Yamahata2016},  so the access transistor itself would not necessarily be the limiting element. AM demodulation imposes a second condition: $f_{\rm mod}$ must lie in the low-frequency range of Fig. 4(b), where the average SN voltage remains close to zero. This range requires $f_{\rm mod}$ to be at least two orders of magnitude smaller than $f_{\rm cutoff}$. Therefore, a GHz-range $f_{\rm mod}$ would require $f_{\rm cutoff}$ far beyond 6.5 GHz , which is not claimed here. The demonstrated regime of $f_{\rm mod}$ is well matched to remote sensing, which benefits from a high carrier frequency while requiring only a modest signal bandwidth. Together with its circuit-level similarity to conventional CMOS image sensors and its applicability to infrared \cite{Nishiguchi2007a} and ultra-low-current sensing \cite{Nishiguchi2024}, this makes the iDGCM a candidate platform for remote sensing with high-frequency carrier signals.

\section{Conclusion}
We demonstrated an all-silicon dual-gate gain-cell memory achieving leakage current below $10^{-22}$ A, inferred from single-electron counting, enabling retention exceeding 1000 s even for an SN capacitance as small as 7.7 aF. Owing to this ultra-low leakage current and the wide tunability of the subthreshold current via the WL voltage, the subthreshold nonlinearity of the access transistor can be exploited via a tunable cutoff frequency, enabling in-memory AM signal demodulation. This in-memory demodulation potentially eliminates external analog front-end circuits, offering a pathway toward ultra-low-power edge-sensing systems.

\bibliographystyle{IEEEtranDOI}
\bibliography{reference}

\end{document}